# An Original Correction Method for Indoor Ultra Wide Band Ranging-based Localisation System

Nezo Ibrahim Fofana, Adrien van den Bossche, Réjane Dalcé, Thierry Val

IRIT, Université Fédérale de Toulouse, CNRS, INPT, UPS, UT1, UT2J
Toulouse-Blagnac, France

`{nezo-ibrahim.fofana - adrien.van-den-bossche - rejane.dalce- val}@irit.fr`

**Abstract.** During this decade, Wireless Sensor Networks (WSNs) brought an increasing interest in the industrial and research world. One of their applications is the indoor localization. The ranging, i.e. the distance evaluation mechanism between nodes, is required to determine the position of the nodes. The research work presented in this article aims to use Ultra Wide Band (UWB) radio links to achieve an efficient ranging, based on Time of Flight (ToF) measurement. A good solution consists in integrating ranging traffic into the usual network messages. However, the ToF ranging process is based on information exchanges which are temporally constrained. Once this information is encapsulated into the usual messages, the temporal constraint cannot be honoured, resulting in important ranging errors due to clock drifts. To mitigate these errors, we have introduced an original dynamic correction technique which enables a precision of twenty centimetres allowing the inclusion of ranging traffic in usual traffic.

## 1  Introduction

The industrial world is a major application area for mobile localisation systems. Spatial and temporal mappings are useful for many applications. GPS is generally used for outdoor localisation, but suffers from a significant power consumption and a reduced performance in indoor environments. Indoor localization and positioning based on other technologies is therefore required. They can take advantage of Wireless Sensor Networks (WSNs) and other wireless communication systems increasingly used in the industrial domain for the exchange of data from the sensors. Localization based on range-free methods, as illustrated by the well-known DV-hop algorithm [1, 2] is simple to implement and is based on hypotheses of cellular connectivity between nodes. However, it is not very accurate. Range-based techniques can improve the localization accuracy. Most wireless nodes can provide an indication of the power level of the received frames, but this method suffers from several drawbacks [3] and is not reliable because of the antenna's polarization and the existence of Non Line Of Sight (NLOS) paths. One of the best solutions is to rely on the Time of Flight (ToF) [4] between sender node and receiver node. This technique requires a precise timestamping of signal at the physical layer, which is made possible through a radio transmission technology like Ultra-Wide Band (UWB) or even Chirp Spread Spectrum (CSS).The DWM1000 [5] base on UWB physical layer of the IEEE 802.15.4 standard is a very powerful solution. It offers some physical tools to measure the ToF with very high precision, thanks to R-MARKER bit (Ranging Marker) in the frame and picoseconds-level timers included in the transceiver.

While the IEEE 802.15.4 standard offers a normalization of message exchanges between nodes to achieve the ToF measurement such as TWR (Two Way Ranging), but it does not specify how to integrate the ranging service into the protocol stack. Nevertheless, we see clearly see the benefits of a fully integrated ranging protocol in data exchange, that is to say providing the ability to perform a distance measurement between two nodes when these nodes exchange information without ranging-dedicated frames. Our objective is to design an "*opportunistic ranging*", where ranging messages are encapsulated into usual messages (data, acknowledgment…). The final objective of our work is to provide such a service, included in the protocol stack, as transparent as possible, minimizing protocol overhead, while staying accurate on the distance measurement. To achieve this service, one of the difficulties is due to the time constrained characteristic of ranging messages. Messages will be delayed, because of the Medium Access Control (MAC) process, introducing discrepancies because the nodes' transceivers use different crystals, which introduce ranging errors. This paper deals with this last point, introducing an original correction method for the ranging system.

The remainder of the article is organized as follows: after this introduction, we begin with a brief state of the art of ranging techniques and testbeds. Then we re-examined the well-known problem of clock drift, because the delay in ranging message exchanges is an important aspect of our problem; We then present our measurements performed on a real testbed that confirm the reality of the problem. Next, we introduce our original correction method and evaluate its performance on the same testbed, before concluding and presenting our perspectives.

## 2  Related Works

Many algorithms and localization systems have been developed for indoor WSNs. The algorithms proposed in the literature can be classified into two categories: the *Range-free* type and the *Range-based* type. The *Range-free* family uses the connectivity and the number of hops in order to estimate the node's position. The second category is based on measurements between nodes that can be converted into distance or angle. This type of localization proceeds in two phases: the ranging phase and the position calculation phase. In the ranging phase, the information is either the Time Difference of Arrival (TDOA) [6], the Angle of Arrival (AOA) [7] or the Received

Signal Strength Indicator (RSSI) [3] to obtain the distance between two nodes. With this information, each node can calculate its own 3D-coordinates using classical localisation algorithms such as trilateration or triangulation. The RSSI technique uses signal propagation in order to convert the received signal strength in distance. It doesn't require additional hardware because transceivers are usually equipped with electronic circuits allowing to associate each received frame to a power level. However, this technique suffers [3] usually of several drawbacks related to high variations in radio propagation, the mobility of nodes, or even antennas directivity. The evaluation of distance between two nodes is altered, and the resulting location is then less precise. Ranging accuracy can be improved to respond to requirements of the location. We achieve this by proposing a dynamic correction exploiting the protocols based on ToF and using the properties of the UWB physical layer of the IEEE 802.15.4 standard: UWB [8], because the RMARKER bit is not available in DSSS. This bit in the header of UWB physical layer (figure 1) allows to timestamps with a great accuracy a moment in a ranging frame (figure 2). The RMARKER bit serves as reference for frame timestamping. The existing UWB transceivers are capable of producing a picoseconds-level timestamp for both incoming and outgoing frame.

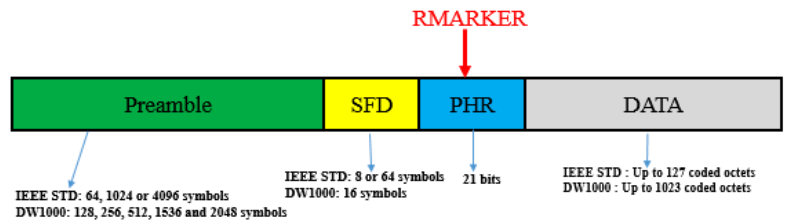

**Fig. 1.** Frame format: IEEE STD and DWM1000 compliant

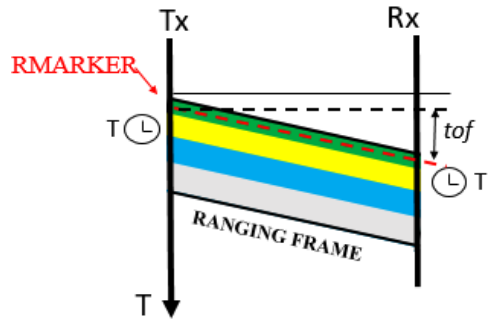

**Fig. 2.** RMARKER in ranging frame

In fact, temporal information provided by nodes in a UWB network offer is more ranging reliable [9] than data from the received power level. Ranging methods based on the ToF use a set of messages to determine the ToF. During transmission and reception of frames, nodes timestamp those moments on the physical layer. These timestamps are used to calculate the ToF and therefore the distance between transmitter and receiver. Multiple protocol variants are possible and are presented in the following sections.

### 2.1 Ranging protocols

**TWR (Two Way Ranging).** TWR consists in a sequence of three messages (Figure 3a). As indicated before, nodes A and B timestamp both outgoing and incoming frames. Node A extracts the timestamps collected by B from DATA_REPLY frame and applies the following equation.

$$ToF_{TWR} = \frac{t_4 - t_1 - (t_3 - t_2)}{2} \tag{1}$$

While it is simple to implement, TWR is also very sensitive to timing inaccuracies due to clocks drift.

**SDS-TWR (Symmetric Double-Sided Two-Way Ranging).** In order to compensate the weakness of TWR, [10] proposes to symmetrise the protocol by adding a message to TWR basic ranging. This symmetry allows SDS-TWR (figure 3b) to reduce the effect of differences in the clocks of two nodes involved, in turn reducing the ranging error. The ToF is computed with equation (2) in this case:

$$ToF_{SDS-TWR} = \frac{t_4 - t_1 - (t_3 - t_2) + t_6 - t_3 - (t_5 - t_4)}{4} \tag{2}$$

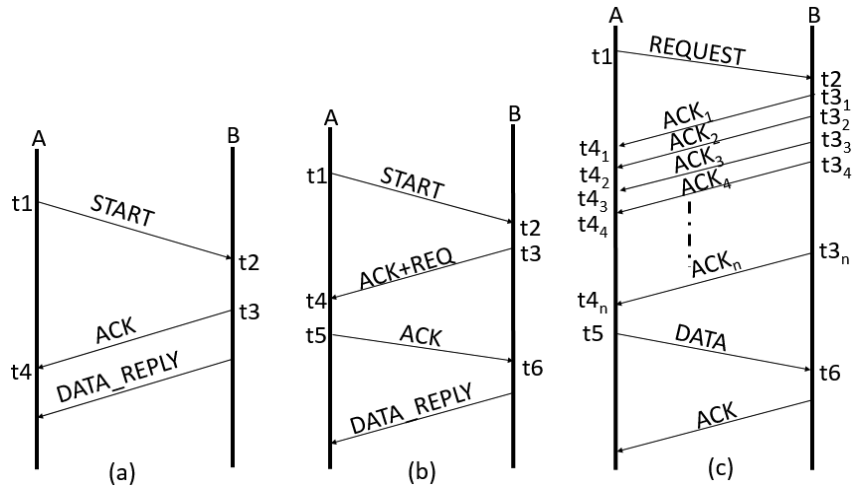

**Fig. 3.** A ranging with TWR (a), SDS-TWR (b) and SDS-TWR-MA (c) protocols.

**SDS-TWR-MA (Symmetric Double-Sided Two-Way Ranging Multiple Acknowledgements).** The SDS-TWR-MA was introduced in [11]. The purpose of this protocol is to reduce the inaccuracy in the original SDS-TWR. Since the clock difference varies over time, many estimates of distance are collected and averaged to reduce the impact of random disturbances. To obtain these "k" estimates, SDS-TWR must be executed "k" times. The multiple execution was used in [12]. In order to reduce the overhead due to multiple execution, the authors of have developed a version of SDS-TWR with multiple acknowledgments called SDS-TWR-MA, where the number of frames, therefore time markers is increased. For each frame sent, nodes timestamp the arrival time of the message and the arrival time of acknowledgment. At the end of the exchange, a last frame containing all the timestamps is sent to the node which initiated the protocol. This solution however has the disadvantage to have a large number of dedicated messages.

Our objective is to finely characterize these different protocols, and to propose an improvement in order to incorporate ranging traffic in the usual network traffic such as data, ack and beacon messages. This work had advantageously used a hardware and software platform developed by our team.

### 2.2 Testbeds for evaluating protocols performance in real conditions

Some research projects have helped to design and deploy real open environments, open and accessible, allowing users to evaluate the protocols performance in real environment. The term testbed is commonly used to refer to these platforms, which may include a large number of nodes, allowing users to test their protocols on real nodes considering the scale factor.

One of the first public large-scale testbeds in France is the SensLab project (INRIA). An important step towards the Internet of Things was the deployment of the FIT/IoT-LAB [13] platform in 2012. One of its rivals, in terms of size, is the SmartSantander platform: each of these testbeds hosts several thousands of nodes.

While most mentioned testbeds are based on the physical layer specified in IEEE 802.15.4-2006, 868 MHz and 2.4 GHz, we found a few public testbeds which propose less common modes of transmission such as UWB. Yet today, many ambitious projects base on this physical layer like the project LoRA Fabian [14] or Freescale [15] for example. The OpenWiNo platform is used in the context of this article. A presentation of this testbed will be provided in section 4. The version [5] used in this study implements the IEEE 802.15.4 UWB specification.

## 3 Problem

Ranging protocols previously presented suffer from a common problem: they all need dedicated messages for ranging. Especially, those exchanges should be performed on the shortest possible duration, which greatly constrains the message scheduling which may impact the Medium Access Control layer. Moreover, the ToF estimation is based on the use of two clocks, one on each node, assumed to be identical. TWR protocol (equation 1) considers that timestamps $t_2$ and $t_3$ are based on the same crystal as $t_1$ and

$t_4$, whereas on a real system, the two nodes are distinct. There is a difference between clocks, which introduces a bias in the ToF since $\{t_1, t_4\}$ and $\{t_2, t_3\}$ are not based on the same clock source. We will present an error model which focuses on this problem later in this article. If we increase the inter-message delay, this difference introduces a bias in the ToF, because of the accumulation of the drift's impact. SDS-TWR, imposing symmetry in the ranging protocol, precisely compensates the error introduced by the clock difference. However, this symmetry is made possible by an additional message, which increases the solution's cost (overhead). As such, if one of the two protocols should be the basis for a work of integration into an existing traffic, TWR would be less constrained because it requires only two messages for the materialization of the timestamps $t_1$, $t_2$, $t_3$ and $t_4$, followed by a message containing $t_2$ and $t_3$ sent to the node at the initiative of ranging.

In order to integrate ranging traffic in the usual network traffic (data, beacons, ack, etc.), it is necessary to evaluate the real impact of clock drift on ranging measurement. In fact, network traffic will be encapsulated in messages which will be more or less delayed depending on the MAC protocols. In the case of CSMA/CA, the *backoff* significantly delays the transmission of messages. If ranging traffic is included in a series of beacons, the delay between beacons may be several tens of milliseconds. The real impact of this delay on the performance of ranging must be evaluated, preliminary to our work.

Finally, if SDS-TWR introduces a symmetry in order to compensate drift related errors introduced by the drift, this symmetry implies a fine control of transmission time: it seems impossible, in the context of an encapsulation of ranging traffic in the usual network traffic, to maintain this symmetry, since transmission instants will depend on the MAC protocols, and medium access method. A non-symmetrical approach as TWR is therefore preferred.

## 4     Preliminary measurements based on real physical nodes

In contrast to many studies which are based on theoretical assumptions, formal calculations, and more or less macroscopic simulations, our goal here is to finely characterize the real world based on real physical nodes. We achieve this through our fast prototyping platform [16] for protocols OpenWiNo. In fact, few public testbeds today are implementing UWB radios. We first implemented the well-known protocols TWR and SDS-TWR, to validate our environment and, on the other hand, to compare their performance. We used UWB transceivers developed by DecaWave [17]. Secondly, we evaluated the impact of the clock drift through experiments.

**Testbed description.** OpenWiNo is an open source protocol development environment for WSN and the device layer of the IoT. It allows fast prototyping of original protocols, in C-language, for execution on a testbed of real nodes called "WiNos" (Wireless Node). The WiNos are developed using an Open Hardware approach. This allows a great versatility on the hardware; it is very simple, for example, to change a WiNo's physical layer: it is only a matter of replacing the transceiver and associated library (Table 1). In this study, we used the DecaWiNo [5], a WiNo built on UWB transceivers developed by DecaWave, compliant with the IEEE 805.15.4 2011 standard, and an ARM CORTEX M4 (Teensy 3.1 Arduino-compliant board: Freescale MK20DX256VLH7 rated at 72Mhz; 64kb RAM and 256kb flash memories6). In order to use the physical layer efficiently, a dedicated library has been designed research team: this software will soon be made available to the scientific community.

**Table 1.** Features of developed WiNos

|  | **WiNoRF22** | **TeensyWiNo** | **DecaWiNo** | **WiNoVW** |
|---|---|---|---|---|
| **CPU/RAM/ Flash** | ARM Cortex M4 (32bit) 72MHz, 64kB RAM, 256kB Flash (PJRC Teensy 3.1) | | | |
| **Transceiver (library Arduino)** | HopeRF RFM22b : 200-900MHz, 1-125kbps, GFSK/FSK/OOK, +20dBm RadioHead | | DWM1000 UWB IEEE 802.15.4 DecaDuino | Various VirtualWire |
| **Sensors** | temperature, light | idem + pressure, acceleration, compass, gyroscope | temperature, light | |
| **Usage** | WSN, IoT | | IoT with ranging, indoor localization | Very Low rate on medium |
| **Availability** | DIY | snootlab.com | DIY (Do It Yourself) | |

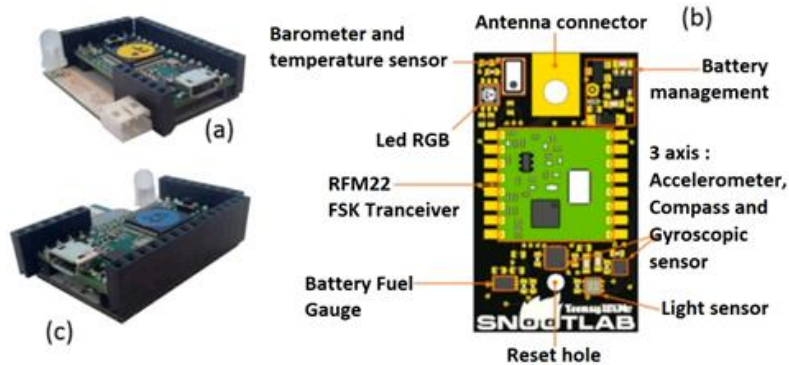

**Fig. 4.** The WiNos nodes

Figure 4 shows three types of WiNo(c): WiNoRF22 (a) TeensyWiNo (b) and DecaWiNo (c). The WiNos are integrated into the Arduino ecosystem, which allow the researcher to easily add hardware components and / or software (sensors and actuators, advanced processing algorithms, interaction devices, library...) in order to prototype a complete solutions.

### 4.1 Implementation and comparison of TWR and SDS-TWR

In a first step, in order to have real temporal characteristics, we performed a comparative metrology between TWR and SDS-TWR on our testbed. This preliminary experiment aims to validate the implementation of hardware and software components, but also to share the raw results from our testbed with the scientific community: the provided data may for example used to develop models of nodes desynchronisation over time or test synchronization solutions; as follows: the maximum distance between the nodes is 5 meters (Figure 5 and 6). Samples are collected every 0.5 m. The measurement protocol is as follows: a new ranging session starts every 200 ms and runs for about 30 s. The measurements were performed with a positioning of the nodes placed 15 cm above the aluminum rail with the PVC bracket. Through experiments, we verified that the impact of the rail on the ranging quality is negligible. In all experiments of the paper, we consider that the nodes are in Line Of Sight (LOS), i.e. without obstacle. The two nodes are powered with 1.8m (6') USB cables.

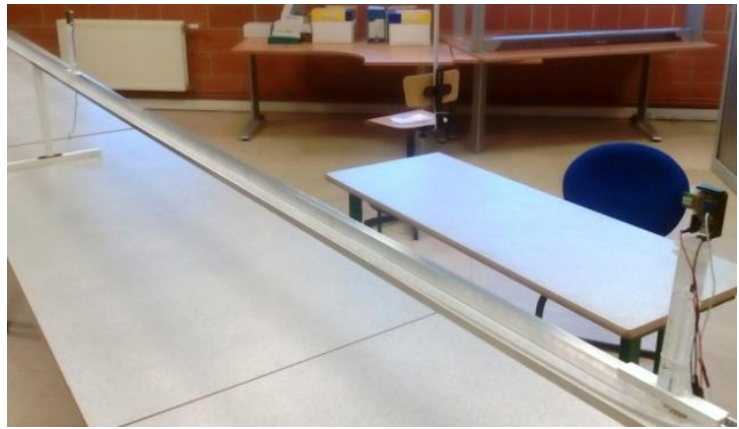

**Fig. 5.** Context of measurements with TWR and SDS-TWR protocols.

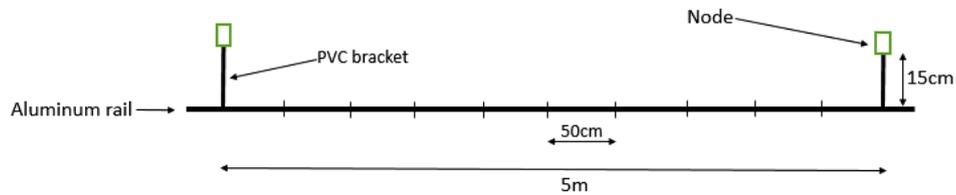

**Fig. 6.** Context of distance measurements

**Ranging with TWR.** The ranging data obtained by the TWR protocol is available in table 2 and represented in figure 7. As we can see, the distance measured in TWR is very close to the actual distance: the absolute error is limited (15cm) and the low standard deviation shows that the results are reproducible. Moreover, the results match the specification provided by the transceiver's manufacturer, which indicates an accuracy of ten cen. This step allows us to conclude positively about the proper implementation of TWR protocol in our environment.

**Table 2.** Error Summary in TWR (in meters)

| Actual distance | Dist. TWR | average error | max error | min error | standard deviation |
|---|---|---|---|---|---|
| 0,5 | 0,339 | -0,160 | -0,11 | -0,24 | 0,029745754 |
| 1 | 0,881 | -0,118 | -0,05 | -0,17 | 0,02427616 |
| 1,5 | 1,417 | -0,082 | -0,03 | -0,14 | 0,023683855 |
| 2 | 1,993 | -0,006 | 0,05 | -0,06 | 0,024150096 |
| 2,5 | 2,480 | -0,019 | 0,03 | -0,07 | 0,017567371 |
| 3,5 | 3,481 | -0,018 | 0,03 | -0,07 | 0,022417601 |
| 5 | 5,046 | 0,046 | 0,08 | 0 | 0,016043844 |

**Ranging with SDS-TWR.** The ranging data is available in table 3 and represented in figure 7. The experimental conditions are the same as those presented previously with the TWR protocol. We make the same observation, the distance measured by SDS-TWR is very close to the actual distance, which confirms here too the good implementation of the protocol in our environment.

**Table 3.** Error Summary in SDS-TWR (in meters)

| Actual distance | Dist. SDS_TWR | average error | max error | min error | standard deviation |
|---|---|---|---|---|---|
| 0,5 | 0,4 | -0,099 | -0,05 | -0,15 | 0,02623868 |
| 1 | 0,942 | -0,057 | -0,02 | -0,09 | 0,01638611 |
| 1,5 | 1,482 | -0,017 | 0,03 | -0,07 | 0,02390231 |
| 2 | 2,061 | 0,061 | 0,11 | 0,01 | 0,02122458 |
| 2,5 | 2,546 | 0,046 | 0,08 | 0,01 | 0,01724975 |
| 3 | 3,067 | 0,067 | 0,11 | 0,03 | 0,02023639 |
| 5 | 5,107 | 0,107 | 0,15 | 0,07 | 0,01559689 |

### 4.2 Discussion

The experiments showed that the error on both SDS-TWR and TWR does not exceed 15 cm and has a low standard deviation (~ 0.02). This implies a high reproducibility of the experiments: therefore, it is not necessary a hundred measurements per distance to obtain accurate results. This is an encouraging trait since high variability would have required many executions of the ranging protocol in order to stabilize the distance estimation and would have hampered the encapsulation of the ranging protocol in usual network traffic. This characteristic will allow the reduction of the number of measurements.

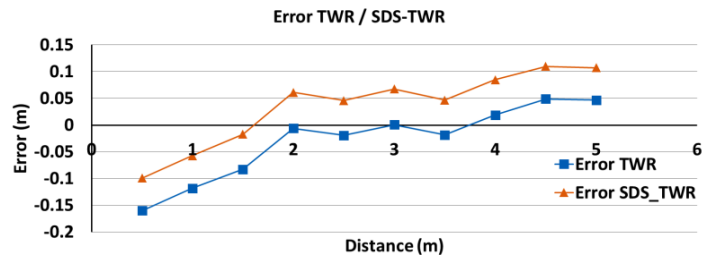

**Fig. 7.** Representation of SDS-TWR / TWR error as a function of the distance

We also noticed that there was little difference between TWR and SDS-TWR in terms of ranging error, although SDS-TWR has been designed to reduce the errors related to clock differences in TWR. The difference between the two curves of figure7 appears as is it could be compensated by a simple calibration. The reason for this resemblance could be the very short duration of the frame exchanges: the protocol duration being low, the impact of the drift remains limited. In both cases, the short distances are impacted by a negative error, while the highest are impacted by a positive error. Again, a calibration is possible.

This preliminary phase shows that the asymmetric approach of TWR protocol presents quite good results and can be retained as a candidate in the rest of our work, to study its incorporation into the usual network traffic.

## 5  Artificial delays and TWR performance

In this section, we will evaluate the ranging accuracy in the presence of artificial delays introduced between the START and ACK messages in TWR (Figure 8). In fact, when the ranging messages will be encapsulated in the usual traffic of the network, unavoidable delays will be introduced in particular by the medium access method. Therefore, we will investigate the consequences on the accuracy of a delay introduced in TWR.

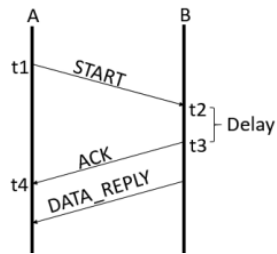

**Fig. 8.** The TWR protocol with a delay between exchanges

In order to get closer to the conditions of a real communication, we introduced an artificial delay between $t_2$ and $t_3$ on node B and increased the timeout value on node A. This will allow the difference between the clocks to increase, degrading the ranging performance.

A new experiment was performed on 11 positions regularly spaced 0.5 meter between nodes A and B (from 0.5 m to 5.5 m). For each point, we set a delay of 1, 2, 3, 4, 5, 6, 7, 8, 9, 10, 16 and 21 ms. Thirty measurements were performed for each delay value, yielding 240 samples per position. A new ranging session begins every 200ms. The results for inter-node distance of 2m are presented on figure 10.

We can observe the impact of this variable delay on the accuracy of the ranging: the performance deteriorates with the increase in the delay. Large delays also imply that the nodes stay in the RX state longer: since reception is very energy-consuming [17], this leads to an increase in the transceiver's temperature. This im-pacts the resonance frequency of the quartz ad further degrades performance. Figure 9 has been captured using a laser thermometer and shows that, the longer the node stays in reception, the higher the temperature. While it is possible to alter the protocol in order to avoid this situation, it remains a general case which must be taken into account.

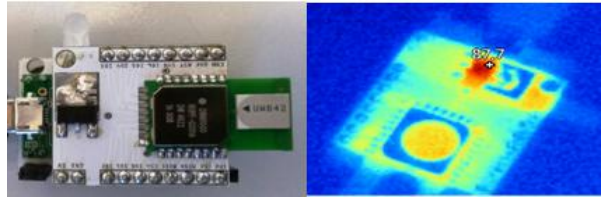

**Fig. 9.** Node DW1000: image in the visible and infrared

## 6   Proposed error mitigation method

In order to compensate the error introduced by the increase of the delays in TWR, we propose a correction method based on the measurement of the clocks' relative drifts.

### 6.1   Principle of the correction

Let's consider nodes A and B, A being the initiator of the ranging protocol. The equation of ToF in TWR was presented in equation 1. In reality, $t_4$-$t_1$ and $t_3$-$t_2$ are periods characterized by different clocks, clocked at frequencies $f_A$ and $f_B$. On the assumption that $f_A = f_B$, ToF estimation is correct. However, in reality, the clocks will not be clocked at the same speed. Let's rewrite the equation while taking into account the fact that $f_A$ does not equal $f_B$. With k the coefficient differentiating these two clocks defined as:

$k = \frac{f_B}{f_A}; k \approx 1$.

The general formula of ToF becomes:

$$ToF_{Corrected} = \frac{t_4 - t_1 - k * (t_3 - t_2)}{2} \qquad (3)$$

This allows compensating the error introduced by the drift of two clocks, especially when the time between $t_2$ and $t_3$ is important.

The DW1000 on which our testbed is based features a *ClockOffset* functionality: by analyzing the corrections made by the PLL to decode the signal, it provides an estimate of the difference between the transmitting and receiving clocks, expressed in parts per million (ppm). Through the DecaDuino library, the protocols in the upper layer have access to this information. In our case, node A applies the correction as formulated in equation 4. This single measurement allows the mitigation of both drift and temperature related errors.

$$ToF_{Corrected} = \frac{t_4 - t_1 - (1 + ClockOffset * 10^{-6})(t_3 - t_2)}{2} \qquad (4)$$

### 6.2 Evaluation

In the last round of measurements, the *ClockOffset* if used to enhance ranging performance. The experiment is taking place in the same conditions as previously. Figure 10 shows the absolute error before (in red) and after correction is applied, as a function of the delay. These results correspond to an inter-node distance of 2m. Table 4 summarizes all results.

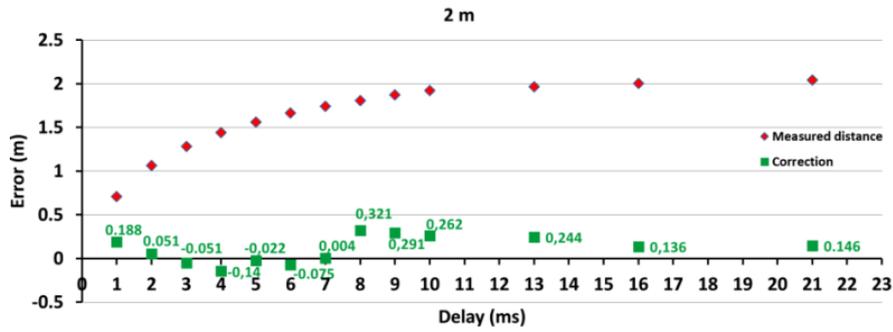

**Fig. 10.** Representation of the error and correction as a function of delay

**Table 4.** Summary table of distances corrected.

| Distance \ Delay | 0,5 m | 1,5 m | 2,5 m | 3,5 m | 4,5 m | 5,5 m |
|---|---|---|---|---|---|---|
| 1 ms  | -0,26 | 0,06  | 0,15  | 0,06  | 0,24 | 0,23  |
| 2 ms  | -0,32 | -0,04 | -0,01 | 0,04  | 0,21 | 0,31  |
| 3 ms  | -0,29 | -0,07 | 0,05  | 0,00  | 0,24 | 0,27  |
| 4 ms  | -0,37 | -0,02 | 0,17  | 0,02  | 0,22 | 0,38  |
| 5 ms  | -0,17 | -0,09 | 0,05  | -0,10 | 0,36 | 0,17  |
| 6 ms  | -0,30 | -0,12 | 0,11  | 0,11  | 0,34 | 0,23  |
| 7 ms  | -0,27 | -0,04 | 0,03  | 0,05  | 0,04 | 0,24  |
| 8 ms  | -0,23 | -0,12 | 0,19  | 0,05  | 0,16 | -0,01 |
| 13 ms | -0,16 | 0,06  | 0,24  | 0,12  | 0,32 | 0,06  |
| 16 ms | -0,11 | 0,04  | 0,27  | 0,18  | 0,38 | -0,02 |
| 21 ms | -0,17 | 0,20  | 0,19  | 0,22  | 0,35 | 0,28  |

It is worth signaling that the ranging error has decreased considerably once the correction was applied: for delays up to 7°ms, the ranging error remains under 20cm, which brings us back to the values initially measured for the time constrained version of the protocol. The error mitigation scheme we propose allows the same performance while relaxing the time constraint. This makes the inclusion of the ranging traffic in the network's communication flow a feasible task.

## 7      Conclusion

In this article, we addressed a fundamental issue in the field of WSN-based localisation. Improving the ranging performance will allow the creation of innovative services in the IoT context. We introduced the context of our work and the benefits of using TOF compared to RSSI. We introduced different ranging protocols encountered in the literature. In order to offer an integrated localisation service in the protocol stack, the ranging protocols should be integrated into the usual traffic (data, ACK, beacons). For this, they need to be tolerant-delays, which is not the case in theory: this sensitivity to delay was introduced in section 3 and confirmed through measurements on a real testbed in section 5. In order to reach our goal, we introduced an original and efficient correction of ranging error, based on clock drift estimation. We proposed an implementation of this correction on our testbed and evaluated the correction pragmatically. The results are satisfying and show that after correction, the ranging messages can be exchanged without any temporal constraint, with an error of about twenty centimeters in the worst case. Therefore, the ranging protocol will be integrated in the common messages in the network.

Following the results, it appears already some indications that we give to our work. Several ways are being explored. In the near future, a ranging protocol based on the cost-effective use of beacons and data frames will be investigated. As a first step, an improvement of TWR will be proposed to allow a continuous localization, in background, of several neighbors to one hop in a multi-hop topology, in order to achieve the final objective and have a complete localization system included in the protocol stack. Another perspective is to improve the ranging in Non Line-Of-Sight (NLOS) situations: we will have the opportunity to evaluate the performance of the UWB technology which is said to be robust to multipath. Finally, one last perspective concerns the testbed used and its opening to the public we plan to begin by making a set of 10 nodes remotely available to fellow researchers through a web interface.